\documentclass[twocolumn,unsortedaddress, aps]{revtex4-1}
\usepackage{graphicx}
\usepackage{amsmath}
\usepackage{graphics}
\usepackage{amsmath}
\usepackage{amsfonts}
\usepackage{amssymb}
\usepackage{xcolor}
\usepackage{subcaption}
\usepackage{subcaption}
\usepackage[bookmarks=false]{hyperref}
\hypersetup{colorlinks=true, citecolor=blue, urlcolor=blue, linkcolor=blue}

\begin{document}



\title{First-passage time theory of  activated rate chemical processes in electronic molecular junctions}

\author{Riley J. Preston}
\affiliation{College of Science and Engineering, James Cook University, Townsville, QLD, 4811, Australia }

\author{Maxim F. Gelin}
\affiliation{School of Sciences, Hangzhou Dianzi University, 310018 Hangzhou, China}

\author{ Daniel S. Kosov}
\affiliation{College of Science and Engineering, James Cook University, Townsville, QLD, 4811, Australia }

\begin{abstract}
Confined nanoscale spaces, electric fields and  tunneling currents make the molecular electronic junction an experimental device for the discovery of  new, out-of-equilibrium chemical reactions.
Reaction-rate theory for current-activated chemical reactions is developed by combining a Keldysh nonequilibrium Green's functions treatment of  electrons, Fokker-Planck description of the reaction coordinate, and Kramers' first-passage time calculations. The NEGF provide an adiabatic potential as well as a diffusion coefficient and temperature with local dependence on the reaction coordinate.  Van  {    Kampen's} Fokker-Planck equation, which describes a Brownian particle  moving in an external potential  in an inhomogeneous medium with a position-dependent friction and diffusion coefficient, is used to obtain an analytic expression for the first-passage time.
The theory is applied to several transport scenarios: a molecular junction with a single, reaction coordinate dependent molecular orbital, and a model diatomic molecular junction. We demonstrate the natural emergence of Landauer's blowtorch effect as a result of the interplay between the configuration dependent viscosity and diffusion coefficients. The resultant localized heating in conjunction with the bond-deformation due to current-induced forces are shown to be the determining factors when considering chemical reaction rates; each of which result from highly tunable parameters within the system.
\end{abstract}

\maketitle

\newpage
\section{INTRODUCTION}

A molecular junction is a single molecule confined in the nanoscale gap between two macroscopic, conducting leads. An applied voltage allows for the flow of electronic current across the system through the valence states of the molecule. Large current densities and power dissipation give way to strong current-induced forces and bond-selective heating which act to destabilize the molecular configuration within the junction, resulting in molecular conformational changes including telegraphic switching \cite{pistolesi10,dundas12,preston2020,kershaw2020}, \  along with providing the necessary energy for total bond rupture\cite{dzhioev11,dzhioev12C,peskin2018,thoss11,li2015,li2016,fuse}.  This is obviously an undesirable feature  for promoting molecular electronics as  a possible avenue for moving beyond the traditional silicon semiconductor technology into a regime of highly efficient and tailorable molecular-scale devices, and so a thorough theoretical understanding is required for further progress.  Nonetheless, it creates an exciting opportunity to explore and produce new chemical reactions by providing a device which traps a single molecule in a confined space of a few nanometers  where the electric field and current are applied locally and selectively \cite{Huang2013,darwish2016,borca2017}.

The adequate and well established theories have been developed for reaction rate calculations  in gas and condensed phases 
	\cite{zwanzig-book,Coffey,Hanggi1990,Melnikov1991,Hochstrasser96,Miller1998,Pollak05,Greiner20}. 
 However, the development of similar theories for molecules in an electronic junction environment is no simple task and as such, the scope of theoretical work is still very limited. Three types of approaches have been proposed to model current-induced dissociation. The first is based on 
the rate equation approach where a single harmonic vibration is pumped beyond the dissociation threshold limit \cite{koch2006,peskin2008}.
The second is a numerically exact scheme, which uses  the hierarchical quantum master equation method in conjunction with a discrete variable representation for the nuclear degrees of freedom to numerically study current-induced dissociation
 \cite{erpenceck2020}.
The third uses Keldysh nonequilibrium Green's functions to obtain a Fokker-Planck equation for the reaction coordinate which is used to compute average escape times and the accompanying reaction rates \cite{dzhioev11,dzhioev12C}. The further development of this approach is the subject of this paper.

Our approach utilizes the Born-Oppenheimer approximation, in which  nuclei within the system are assumed to be slow-moving, classical particles interacting with a sea of fast, quantum electrons. This separation of time-scales enables the use of a Langevin description for the motion of nuclei, in which the forces due to the quantum electrons are conceived through a frictional force which acts to drive the nuclei into a motionless state, a fluctuating force which re-energizes the nuclei, and an adiabatic force which can deform the structure of the reaction potential. This enables the consideration of highly non-trivial behaviour on the nuclear dynamics  at the cost of a fully quantum description. Nevertheless, the method has proven successful in a range of circumstances \cite{subotnik18,subotnik17,subotnik17-prl,Bode11,dzhioev12C,Pistolesi08,pistolesi10,Bode12,kershaw2020,preston2020,fuse,prestonAC2020}.

The method further lends itself to the study of current-induced chemical reaction rates \cite{dzhioev11,dzhioev12C}. 
 The use of Langevin dynamics to compute reaction rates was first explored by Kramers in his seminal 1940 paper \cite{kramers40}, in which the mean first-escape time of a particle trapped in an arbitrary potential well subject to Langevin forces was approximated. 
The next significant step was made in the 1990s, when Kramers' theory was extended to account for position-dependent friction and  generalized Langevin equations describing finite-memory (non-Markovian) fluctuation-dissipation processes \cite{Pollak93,Pollak94,Haynes95,Karplus96}. The effect of a velocity-dependent friction on Kramers' rates was investigated in Ref. \cite{GK2007}.	
However, these studies were limited to the regime of thermodynamic equilibrium. Beyond this regime, the fluctuation-dissipation theorem no longer holds, allowing for the emergence of localized heating effects in analogy to Landauer's proposed blowtorch effect \cite{Landauer1975,Landauer1993}, in which specific configurations of the reaction coordinate may experience heightened temperatures, which may have a significant affect on the evolution of the system. Such systems are not limited to the realms of molecular electronics; the most common examples include numerous  
molecular motors, ratchets, and heat engines \cite{Reimann02,Hanggi09,Cakmak15} as well as various confined nanosystems  \cite{Ray15,Devine17,Rubi17,Holubec17,Rubi20}, notably  of biological significance  \cite{Basak19,Rubi19}.  
Several explicit simulations of Landauer's blowtorch effects in double-well potentials have also  been  performed recently  \cite{Kumar99,Das15,Das19}.

One of the aims of this paper is to further shed light on this topic. A comprehensive understanding of the effects of localized heating on the stability of molecular geometries is required to ensure the productive development of specific functionalities of molecular-scale electronic systems. In this paper, we relax the requirement of thermodynamic equilibrium, allowing for the self-consistent study of the mean first-passage times in model molecular electronic junctions in both the underdamped and overdamped regimes. This is calculated through a Fokker-Planck equation, which arises due to our Langevin description of the reaction coordinate within the junction. To a certain extent  this work is the continuation of two papers of one of the authors (DSK) \cite{dzhioev11,dzhioev12C},
however Ref.\cite{dzhioev11} considered the problem employing the fluctuation-dissipation theorem and Ref.\cite{dzhioev12C} focused on  the underdamped case only.

The paper is structured as follows. In section \ref{Theory}, we demonstrate our calculations for the mean first-passage time in the limiting regimes. This involves the calculation of the current-induced forces in the system, from which a Fokker-Planck description then yields an equation for the mean first-passage times. This is then applied to a simple model of the blowtorch effect in section \ref{blowtorch}. In section \ref{single_results}, we calculate the reaction rates for a single-level junction model, in which current-induced forces are calculated self-consistently within the model. This is then further applied {    to a model  two-level} molecule within the junction in section \ref{multi_results}. Atomic units are used in all calculations such that $\hbar = e = 1$. Boltzmann's constant is set to one $k_B=1$ in all derivations and calculations meaning that the temperature is measured in units of energy.

\section{THEORY}
\label{Theory}

\subsection{Hamiltonian}
\label{Hamiltonian}
We begin with the general Hamiltonian describing a molecular junction as given by 
\begin{equation}
H(t) = {H}_{M}(t) + {H}_{L} + {H}_{R} + {H}_{LM} + {H}_{MR}.
\label{hamiltonian}
\end{equation}
The total system Hamiltonian is partitioned into the following components; the molecular Hamiltonian ${H}_{M}(t)$, the left and right leads Hamiltonians ${H}_{L}$ and ${H}_{R}$, and the leads-molecule coupling Hamiltonians ${H}_{LM}$ and ${H}_{MR}$ which describe the coupling between the electronic states on the central molecule and the left and right leads, respectively.

The molecular Hamiltonian takes the form: 

\begin{equation}
{H}_{M}(t) = \sum_{ij} h_{ij} ( x(t) ) d^{\dag}_{i} d_{j} +\frac{p^2}{2m} + U(x),
\label{molecularhamiltonian}
\end{equation}
where the operators $d^{\dag}_{i}$ and $d_{j}$ denote the creation and annihilation operators for the molecular electronic states whose energies are given by the Hamiltonian matrix elements $h_{ij}(x(t))$. Note the explicit time dependence here, which arises as a result of the time evolution of the classical reaction coordinate $x$. The last two terms in $H_M$ are not quantum mechanical operators, $p^2/2m$ is the kinetic energy for the motion of the reaction coordinate and $U(x)$ is the  classical potential energy.
 
\

The leads Hamiltonian is taken in the standard form of non-interacting electrons reservoirs:

\begin{equation}
{H}_{L} + {H}_{R}  = \sum_{k \alpha} \epsilon_{k \alpha} d^{\dagger}_{k \alpha} d_{k \alpha},
\end{equation}
where the creation and annihilation operators are given by $d^{\dagger}_{k \alpha}$ and $d_{k \alpha}$, and the subscript $k\alpha$ denotes the operator acting on state $k$ in the $\alpha$ lead which has energy $\epsilon_{k \alpha}$.

\

Finally, the system-lead coupling Hamiltonians $H_{LM}$ and $H_{MR}$ are given by:
\begin{equation}
{H}_{LM} + {H}_{MR} = \sum_{k\alpha i} \Big( t_{k \alpha i}  d^{\dagger}_{k \alpha} d_{i} + \text{h.c.} \Big).
\end{equation}
The matrix elements $t_{k \alpha i}$ (and their conjugates) describe the tunnelling amplitudes between lead states $k \alpha$ and the molecular orbitals $i$. 

\subsection{Green's Functions and Self-Energies}
\label{GFandSE}

The foundational components from which we build our model for the classical motion within a molecular junction are the adiabatic Green's functions. The required Green's function components (advanced, retarded, lesser and greater)\cite{negf} can be solved for from the Keldysh-Kadanoff-Baym equations, expressed in the Wigner space by\cite{haug-jauho}
\begin{multline}
\Big(\omega+\frac{i}{2}\partial_t-e^{\frac{1}{2i}\partial_{\omega}^{G}\partial_t^{h}} h(t)\Big)G^{R/A}(t,\omega)=I\\
+e^{-\frac{1}{2i}\partial_{\omega}^{\Sigma}\partial_t^{\mathcal{G}}}\Sigma^{R/A}(\omega)G^{R/A}(t,\omega),\label{eqm4}
\end{multline}
and 
\begin{multline}
\Big(\omega+\frac{i}{2}\partial_t-e^{\frac{1}{2i}\partial_{\omega}^{G}\partial_t^{h}} h(t)\Big)G^{</>}(t,\omega)=\\
e^{-\frac{1}{2i}\partial_{\omega}^{\Sigma}\partial_t^{G}}
\Big(\Sigma^{R}(\omega)G^{</>}(t,\omega)+\Sigma^{</>}(\omega)G^{A}(t,\omega)\Big),\label{eqm3}
\end{multline}
where we have shown the retarded/advanced and lesser/greater equations collectively. Here, we have already assumed that our molecule-leads coupling components $t_{k\alpha i}$ are time-independent, which makes the self-energies only $\omega$-dependent.

In alignment with our previous work\cite{kershaw17,kershaw18,kershaw19,kershaw2020,preston2020,prestonAC2020}, we assume that the classical motion along the reaction coordinate within the system occurs over long time-scales relative to the characteristic electron tunnelling time.
This provides us with the required small parameter to be able to perturbatively solve (\ref{eqm4}) and (\ref{eqm3}) up to the first order in expansion of the exponents with derivatives. Resultantly, the adiabatic  Green's functions in the Wigner space (Green's functions which instantaneously follow the changes in the reaction coordinate) are given by matrices of the form:

\begin{equation}
G^{R/A}(x ,\omega) = (\omega I - h(x) - \Sigma^{R/A})^{-1},
\end{equation}
\begin{equation}
G^{</>}(x ,\omega) = G^R(x ,\omega)  \Sigma^{</>}(\omega) G^A(x ,\omega),
\end{equation}
where $I$ represents the identity operator in the molecular  space. The corresponding self-energy components are given by

\begin{equation}
\Sigma_{\alpha, i j}^R=  -\frac{i}{2} \Gamma_{\alpha, i j},
\end{equation}

\begin{equation}
\Sigma_{\alpha, i j}^A=  \frac{i}{2} \Gamma_{\alpha, i j},
\end{equation}

\begin{equation}
\Sigma_{\alpha, i j}^<(\omega)=  i f_{\alpha} (\omega) \Gamma_{\alpha, i j},
\end{equation}

and

\begin{equation}
\Sigma_{\alpha, i j}^>(\omega)=  -i (1 - f_{\alpha} (\omega)) \Gamma_{\alpha, i j},
\end{equation}
where we have applied the wide-band approximation to the leads, eliminating any $\omega$ dependence from $\Gamma$. As a result, the elements of the level broadening function are given by
\begin{equation}
\Gamma_{\alpha, ij}  = 2 \pi \rho_\alpha t^*_{\alpha i}t_{\alpha j},
\end{equation}
where $\rho_\alpha$ is the constant, energy-independent density of states for lead $\alpha$ and $t_{i\alpha}$ is the tunneling amplitude   $t_{ik\alpha}$ which no longer depends on $k$.

We additionally solve for the first non-adiabatic correction to the Green's functions, which account for the non-zero velocity of the classical coordinate. The Green's function corrections are given by \cite{Bode12}

\begin{equation}
G_{(1)}^{R/A} = \frac{1}{2i} G^{R/A} \Big[G^{R/A}, \partial_t h  \Big]_{-}G^{R/A},
\label{g1A}
\end{equation}

and 

\begin{multline}
G_{(1)}^{<} = G^R \Sigma^< G_{(1)}^A 
\
+ G_{(1)}^R \Sigma^< G^A
\\\
+ \frac{1}{2i} G^R \Big(\partial_T h G^R \partial_{\omega} \Sigma^< + G^< \partial_T h + h.c.\Big)G^A. 
\end{multline}

\subsection{The Langevin Equation}
Under the overarching assumption that the reaction coordinate $x$ along with its corresponding {    momentum} $p$ are classical variables in our approach due to the separation of time-scales within the system, the equation of motion of the reaction coordinate can be expressed in the form of a quasi-classical Langevin equation, in which the classical motion is dictated by quantum mechanical forces. Our Langevin equation is given by
\cite{Bode12,preston2020,Pistolesi08,pistolesi10,dzhioev12C,kershaw2020}

\begin{equation}\label{Lan}
\frac{d p}{d t} =  - \partial_x U(x) + F(x) -  \frac{\xi(x)}{m}  p  + \delta f(t),
\end{equation}
in which we have the external classical potential $U(x)$, an adiabatic force {    $F(x)$} arising due to the occupation of electronic levels within the molecule, the frictional force and its corresponding viscosity coefficient {    $\xi(x)$}, and the stochastic force $\delta f(t)$. In our model, the adiabatic force takes the form

\begin{equation}
F(x) = \frac{i}{2\pi} \int d\omega \text{Tr} \Big\{ \partial_x h G^<\Big\}.
\label{F}
\end{equation}

Similarly, the viscosity coefficient is defined according to\cite{Bode12,dzhioev12C}

\begin{equation}\label{xix}
\frac{\xi(x)}{m}p = -\frac{i}{2\pi} \int d\omega \text{Tr} \Big\{ \partial_x h G_{(1)}^<\Big\}.
\end{equation}

It is assumed that the fluctuating force can be described by a Gaussian process, defined by

\begin{equation}\label{f1}
\langle \delta f(t) \rangle = 0,
\end{equation}

and 

\begin{equation}\label{f2}
\langle \delta f(t) \delta f(t') \rangle = D(x) \delta (t-t'),
\end{equation}
where $D(x)$ is the quantum-mechanically calculated diffusion coefficient which determines the variance in the fluctuating force. In our model, $D(x)$ is calculated according to\cite{Bode12,dzhioev12C}

\begin{equation}\label{Dx}
D(x) = \frac{1}{2\pi} \int d\omega \text{Tr} \Big\{ \partial_x h G^> \partial_x h G^< \Big\}.
\end{equation}

\subsection{Effective potential energy surface}
The key quantity for chemical reaction rate calculations is the effective potential energy surface.  
Here the effective nonequilibrium potential energy surface is defined via the integration of the
 nonequilibrium force in the Langevin equation
 \begin{equation}\label{Ueff}
 U_{\text{eff}}(x) =U(x) - \int_{x_0}^{x} dy \; F(y).
 \end{equation}
{
	We chose $x_0$ in such a way that  $U_{\text{eff}}(x)$ at the bottom of the potential well,  $x_a$,  is zero,  $U_{\text{eff}}(x_a)=0$. 
	This enforces that the energy $E$ in the formulae below can vary in the range from 0 to $\infty$.  
}
Suppose that the effective potential has the "standard" Kramers problem form, in which case we have a minimum at $x=x_a$  (reactant state) and an energy barrier at $x=x_b$, separating the reactant and product state.

Let's call
{
\begin{equation}
U_a= U_{\text{eff}}(x_a)=0
\end{equation}
}
and 
\begin{equation}
U_b= U_{\text{eff}}(x_b).
\end{equation}

\subsection{Calculations of chemical reaction rates}

	\subsubsection{Fokker-Planck equation with reaction-coordinate dependent viscosity and diffusion} 

The starting point for all derivations of the first-passage times is van Kampen's Fokker-Planck equation
{    for the probability density function $\rho(x,p,t)$} 
 which describes
 a Brownian particle of mass $m$, position $x$ and momentum
$p$ moving in the external potential $U_{\text{eff}}(x)$ in an inhomogeneous
medium with position-dependent friction $\xi(x)$ and diffusion coefficient
$D(x)$ \cite{VanKampen1988b}
{   
$$
\partial_{t}\rho(x,p,t)=
$$
\begin{equation}
\left(-\partial_{x}\frac{p}{m}+\partial_{p}U_{\text{eff}}^{^{\prime}}(x)+\left\{ \partial_{p}\frac{p}{m} \xi(x) +  \partial_{p}^{2} D(x) \right\} \right)\rho(x,p,t).\label{FPEa}
\end{equation}
}
 Van Kampen's Fokker-Planck equation (\ref{FPEa})
is internally consistent (for example, it yields energy- and mass-flow
transport coefficients obeying Onsager's reciprocity relations \cite{VanKampen1991})
and can unambiguously be derived from the Langevin equation  (\ref{Lan}) with Gaussian stochastic force obeying 
(\ref{f1}) and (\ref{f2}). 
The It\^o-Stratonovich dilemma  \cite{VanKampen1981} does
not arise in the derivation of van Kampen's Fokker-Planck equation
(\ref{FPEa}), because the viscosity $\xi(x)$ of Eq. (\ref{xix}) and the diffusion coefficient $D(x)$ of Eq. (\ref{Dx}) are momentum-independent.

\subsubsection{Underdamped limit}

In this section, we present our equations for the mean first-passage times from a reaction potential in an inhomogeneous medium for both the underdamped and overdamped limiting regimes. 
Beginning with van Kampen's Fokker-Planck equation (\ref{FPEa})
for our reaction coordinate $x$ and momentum $p$, we can obtain a one-dimensional energy-diffusion equation which is valid in the underdamped
limit (see Appendix \ref{Und} for the derivation):
{   
\begin{equation}
\label{D}\partial _t\rho (E,t)=\partial _ED(E)\rho^{\text{un}}_{\text{st}}(E)\partial _E\left(\rho^{\text{un}}
_{\text{st}}(E)\right)^{-1}\rho (E,t). 
\end{equation}
}
Here, we have introduced the energy-diffusion $D(E)$ as 

\begin{equation}
D(E)=\frac{\nu (E)}{\Omega (E)}, 
\end{equation}

defined as the ratio of the two energy-dependent functions,

{   
\begin{equation}
\label{nu}
\nu (E)= \sqrt{\frac{2}{m}} \int dx \xi(x)T_{\text{eff}}(x)\sqrt{E-U_{\text{eff}}(x)},  
\end{equation}
and
\begin{equation}
\label{Om}
\Omega (E)= \sqrt{\frac {m}{2}} \int \frac{dx}{\sqrt{E-U_{\text{eff}}(x)}}, 
\end{equation}
}

where we have defined the effective temperature $T_{\text{eff}}$ as

\begin{equation}
T_{\text{eff}}(x)  =\frac{D(x)}{2\xi(x)}.
\label{Teff}
\end{equation}

 In Eqs. (\ref{nu}) and (\ref{Om}) and in all other similar expressions, the integration domain corresponds to $E > U_{\text{eff}}(x)$.

The stationary distribution is written as

\begin{equation}
\label{st}\rho^{\text{un}}_{\text{st}}(E)=Z_{\text{un}}^{-1}\Omega (E)\exp \left\{ -\int_0^EdE^{\prime }
\frac{\mu (E^{\prime })}{\nu (E^{\prime })}\right\}, 
\end{equation}

 with normalization constant 
 
 \begin{equation}
 Z_{\text{un}}=\int dE \, \Omega (E)\exp \left\{ -\int_0^EdE^{\prime }
 \frac{\mu (E^{\prime })}{\nu (E^{\prime })}\right\},
 \end{equation}
 
and

{   
\begin{equation}
\label{mu}
\mu (E)= \sqrt{\frac{2}{m}} \int dx\xi (x)\sqrt{E-U_{\text{eff}}(x)}. 
\end{equation}
}

The energy diffusion equation can be used to calculate the mean first-passage time in the underdamped regime as per
{   
\begin{equation}
\label{tau}
\tau =\int_{U_a}^{U_b}dE^{\prime }\frac 1{D(E^{\prime })\rho^{\text{un}}
_{\text{st}}(E^{\prime })}\int_0^{E^{\prime }}dE^{\prime \prime }\rho^{\text{un}}
_{\text{st}}(E^{\prime \prime }).
\end{equation}
}

\subsubsection{Overdamped limit}

In the overdamped limit,  van Kampen's Fokker-Planck equation (\ref{FPEa}) reduces to the diffusion equation  \cite{VanKampen1988b,VanKampen1988a} which can be cast into a form similar to Eq. (\ref{D}):
{   
\begin{equation}
\label{DvK}\partial _t\rho (x,t)=\partial _x \frac{T_{\text{eff}}(x)}{\xi (x)} \rho^{\text{ov}}_{\text{st}}(x)\partial _x
(\rho^{\text{ov}}_{\text{st}}(x))^{-1}\rho (x,t).
\end{equation}
}
The stationary distribution is given by
{
   
\begin{equation}
\label{EqvK}\rho^{\text{ov}}_{\text{st}}(x)=\frac{1}{Z_{\text{ov}}T_{\text{eff}}(x)}\exp \left\{ -\int_0^xdx^{\prime }
\frac{\partial_x U(x') - F(x')}{T_{\text{eff}}(x^{\prime })}\right\},
\end{equation}
where
\begin{equation}
Z_{\text{ov}}= \int_{-\infty}^{+\infty} dx \frac{1}{T_{\text{eff}}(x)}\exp \left\{ -\int_0^xdx^{\prime }
\frac{\partial_x U(x') - F(x')}{T_{\text{eff}}(x^{\prime })}\right\}.
\end{equation}

The first-passage time in the overdamped regime can be thus written as
\begin{equation}
\label{tauvK}\tau =\int_{x_a}^{x_b}dx^{\prime }\frac {\xi (x^{\prime})}{T_{\text{eff}}(x^{\prime })\rho^{\text{ov}}
_{\text{st}}(x^{\prime })}\int_{-\infty }^{x^{\prime }}dx^{\prime \prime }\rho^{\text{ov}}
_{\text{st}}(x^{\prime \prime }).
\end{equation}
}

To summarize the derivations in section {\ref{Theory}: the main result is the combined use of the adiabatic force (\ref{F}), the position-dependent viscosity (\ref{xix}) and the position-dependent diffusion coefficient (\ref{Dx}) - all obtained from nonequilibrium Green's function calculations - with exact (in underdamped and overdamped limits) expressions (\ref{tau},\ref{tauvK}) for the mean first-passage time $\tau$. The rates for corresponding chemical reactions can obtained as inverse of the first-passage time $1/\tau$.

\section{RESULTS}
\label{Results}

\subsection{The blowtorch effect}
\label{blowtorch}

In this section we investigate Landauer's proposed blowtorch effect\cite{Landauer1993}, in which a non-equilibrium system allows for coordinate-dependent variations to the dissipative forces acting on a particle, which then has an effect on the properties of the steady-state distribution. Landauer's blowtorch effect plays a critical role in chemical reactions in molecular electronic junctions, therefore we first discuss its general features which will be relevant for our subsequent discussion.
For consistency with Kramers' seminal paper\cite{kramers40} on chemical reaction rates, our analysis is formulated in terms of the mean first escape time from the left minimum of a bistable potential, which we model according to a quartic of the form 

\begin{equation}
U(x)=-ax^2 + bx^4,
\end{equation}

where $x$ is our reaction coordinate. The constants $a$ and $b$ are adjustable parameters which determine the width and depth of the minimum. In all tests in this section, we set $a = 0.04$ and $b = 0.008$, such that $U_b = 0.05$. In addition, the particle always begins its trajectory with zero velocity at the minimum of the reaction potential. 
To begin with, the viscosity $\xi_0$ and diffusion coefficients $D_0$ are set to a constant value over the range of $x$, yielding a constant temperature as determined by the fluctuation-dissipation theorem.

In order to introduce an inhomogeneity into the temperature, a Gaussian spike is applied to the diffusion coefficient locally at position $x_0$,

\begin{equation}
D(x) = D_0 + D_{peak}(x),
\end{equation}
where

\begin{equation}
D_{peak}(x) = D_{m} e^{-\frac{(x-x_0)^2}{\sigma ^2}},
\end{equation}
with adjustable width $\sigma$ and magnitude $D_{m}$ parameters. The effective temperature profile is given by (\ref{Teff}),  the effective temperature at the peak is then given by

\begin{equation}
T_{max} = T_0 (1 + \frac{D_m}{D_0}),
\end{equation}

where 

\begin{equation}
T_0 = \frac{D_0}{2\xi_0}.
\end{equation}

This represents Laundauer's so-called "blowtorch" which heats a small segment of the reaction coordinate, as shown diagramatically in Fig.\ref{blowtorchfig}.

\begin{figure}
\includegraphics[width=0.47\textwidth]{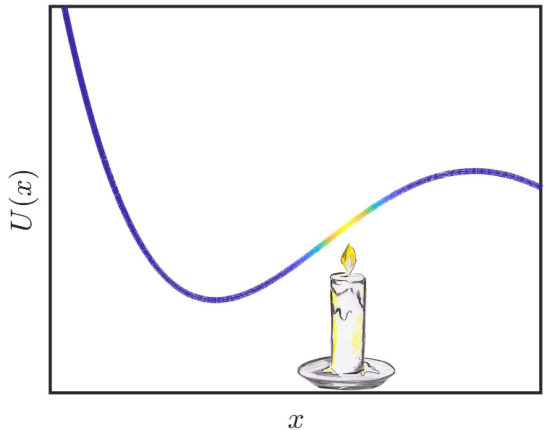}
\caption{An adjustable temperature spike is introduced which heats a chosen part of our reaction coordinate potential.}
\label{blowtorchfig}
\end{figure}

Here, the intention is to study the effect of shifting the temperature spike along the reaction coordinate on the mean first-passage time $\tau$. We analyse the overdamped and underdamped regimes separately for the same parameters except for the  mass $m$ of a Brownian particle, which is chosen to satisfy the desired regime.

\begin{figure}
\centering
\begin{subfigure}{0.47\textwidth}
\centering
\includegraphics[width=1\textwidth]{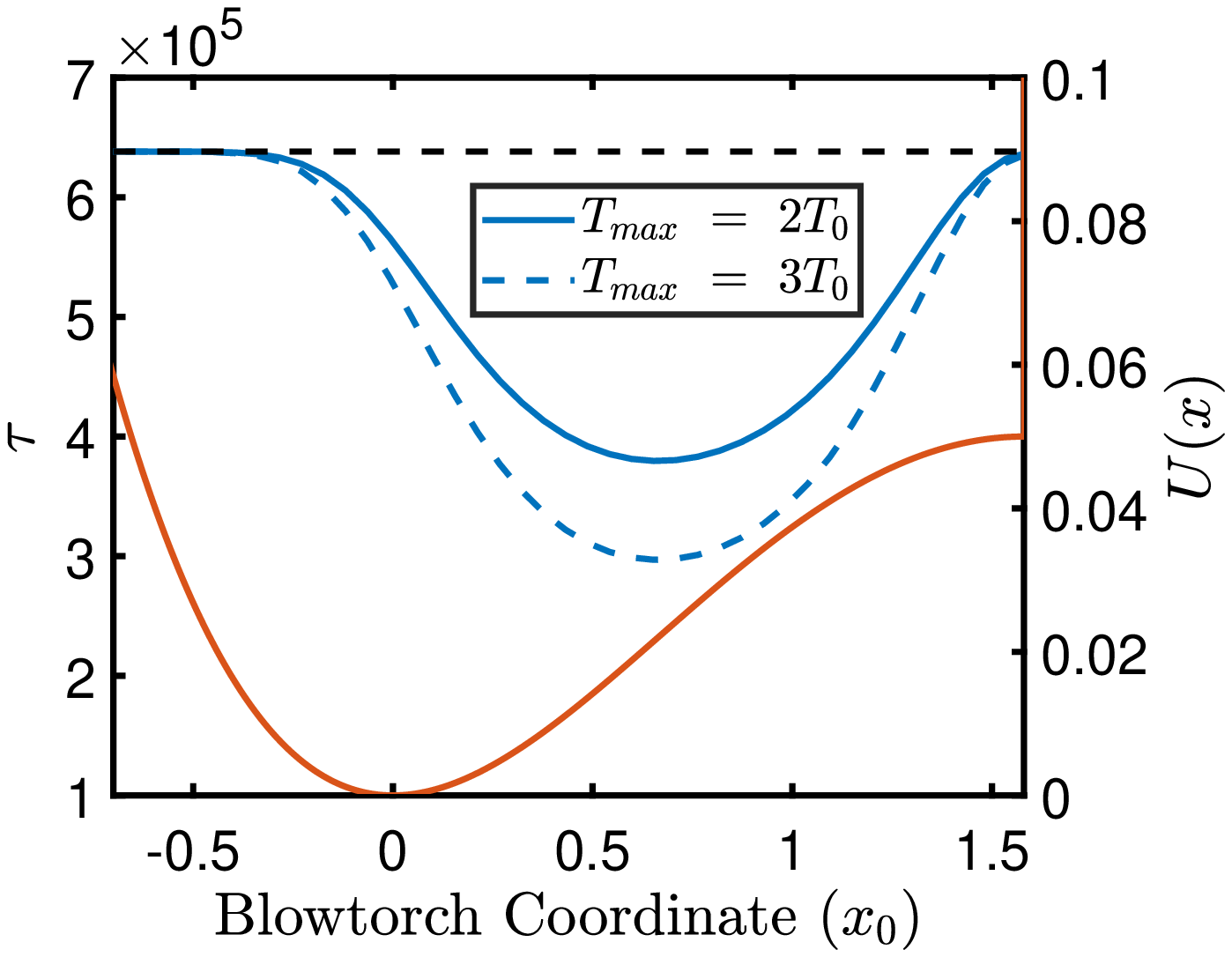}
\caption{}
\label{blowtorch_overdamped}
\end{subfigure}
\begin{subfigure}{0.47\textwidth}
\centering
\includegraphics[width=1\textwidth]{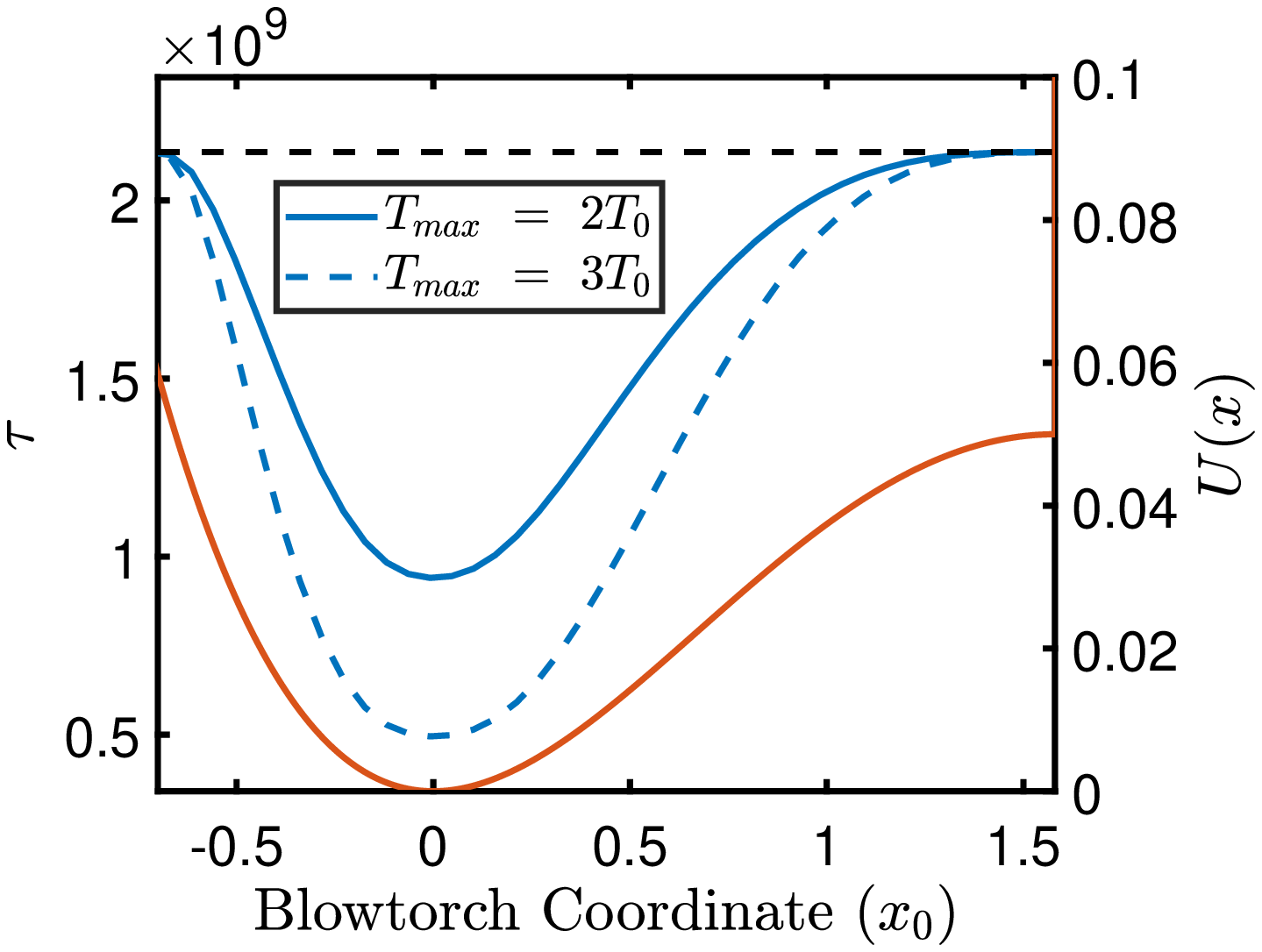}
\caption{}
\label{blowtorch_underdamped}
\end{subfigure}

\caption{The calculated mean first-passage time $\tau$ as a function of the temperature peak's position along the reaction coordinate, for the (a) overdamped and (b) underdamped regime ($m = 1000$a.u). The black line denotes $\tau$ in the absence of an applied blowtorch. Parameters: $D_0 = 0.01$, $\xi_0 = 1$,  $\sigma = 0.05$.}
\label{blowtorch_data}
\end{figure}

In Fig.\ref{blowtorch_data}, we observe the behaviour of the mean first-passage time as the position of the temperature peak is shifted along the reaction coordinate (shown in blue), while the reaction potential is shown as a reference in orange. In the underdamped regime, we observe  $\tau$ to be minimized when the heating is applied to the bottom of the potential. This enables the molecule to heat up quickly at low energies, and repeatedly attain more energy as it passes through this region in a near-harmonic manner.

The overdamped regime differs, in that $\tau$ is minimized when the heating is applied approximately halfway up the potential, around the point of steepest ascent.  In the overdamped regime, the escaping particle will very quickly equilibrate to any given temperature fluctuation to which it is exposed. As such, the heated region causes a flattening of the probability distribution in that region, nullifying the dependence of the distribution on the reaction coordinate. This causes an effective reduction to the height of the energy barrier $U_b$ as elucidated by Landauer\cite{Landauer1975, Landauer1993}; a phenomenon which is maximized when the heating is applied in the region of steepest ascent. We note the counter-intuitive observation that if the heating is applied to the bottom of the potential in the overdamped case, this causes only a small reduction to $\tau$. This is because the particle will quickly lose the obtained energy as it returns to the cooler regions when it attempts to escape.

It is insightful for us to also study the effect of the strength of interaction of a Brownian particle with the environment, while maintaining a homogeneous temperature. This entails that any changes in the diffusion coefficient as a function of $x$ will be counteracted by a corresponding changes in the viscosity coefficient at the same $x$, enforcing a homogeneous temperature as per the fluctuation-dissipation theorem. Here, we perform a similar analysis as above, such that we have a moveable peak of increased interaction (simultaneously locally increased diffusion \textit{and} viscosity) while the temperature is homogeneous. The results of this are displayed in Fig.\ref{interaction_data}.

\begin{figure}
\centering
\begin{subfigure}{0.47\textwidth}
\centering
\includegraphics[width=1\textwidth]{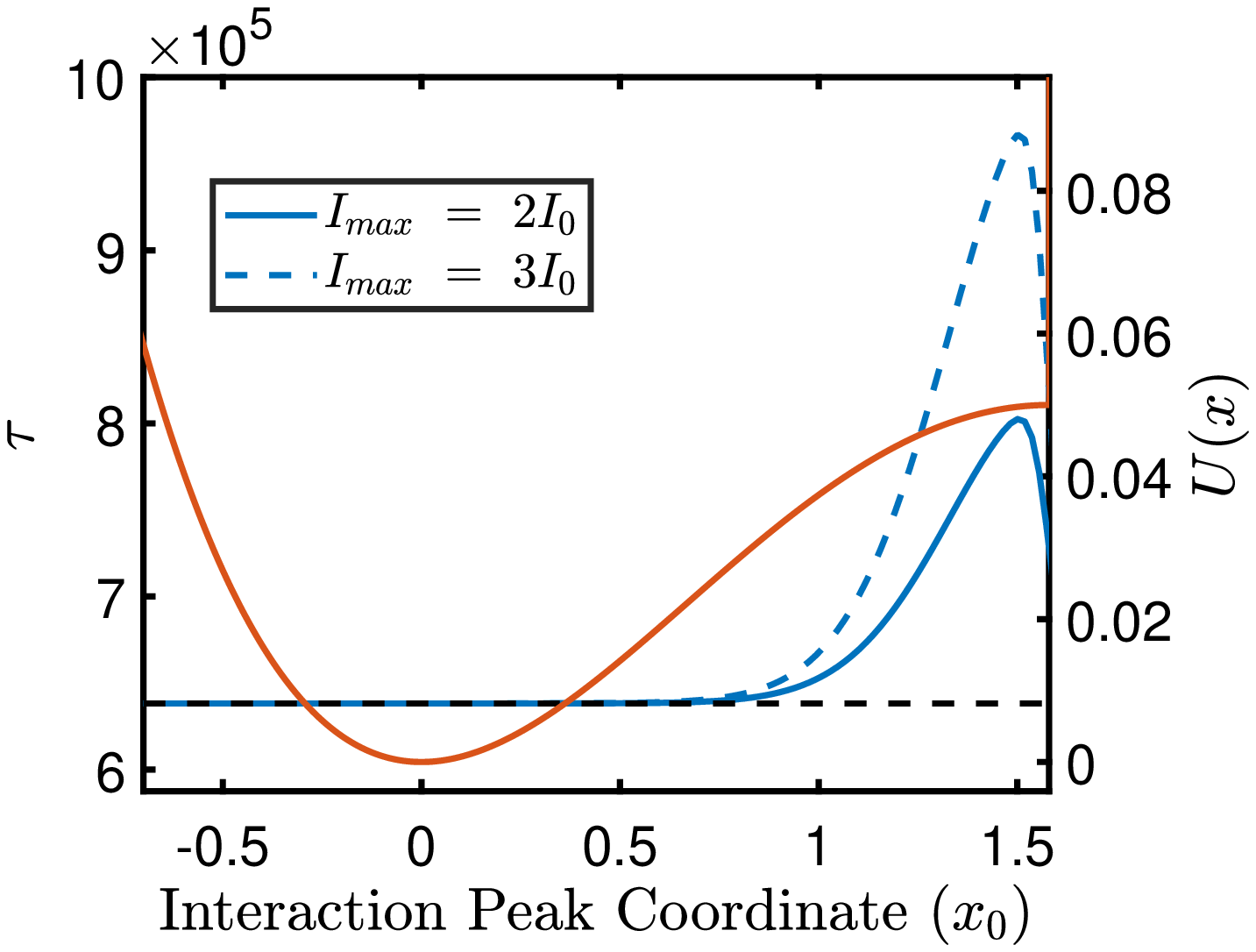}
\caption{}
\label{interaction_overdamped}
\end{subfigure}
\begin{subfigure}{0.47\textwidth}
\centering
\includegraphics[width=1\textwidth]{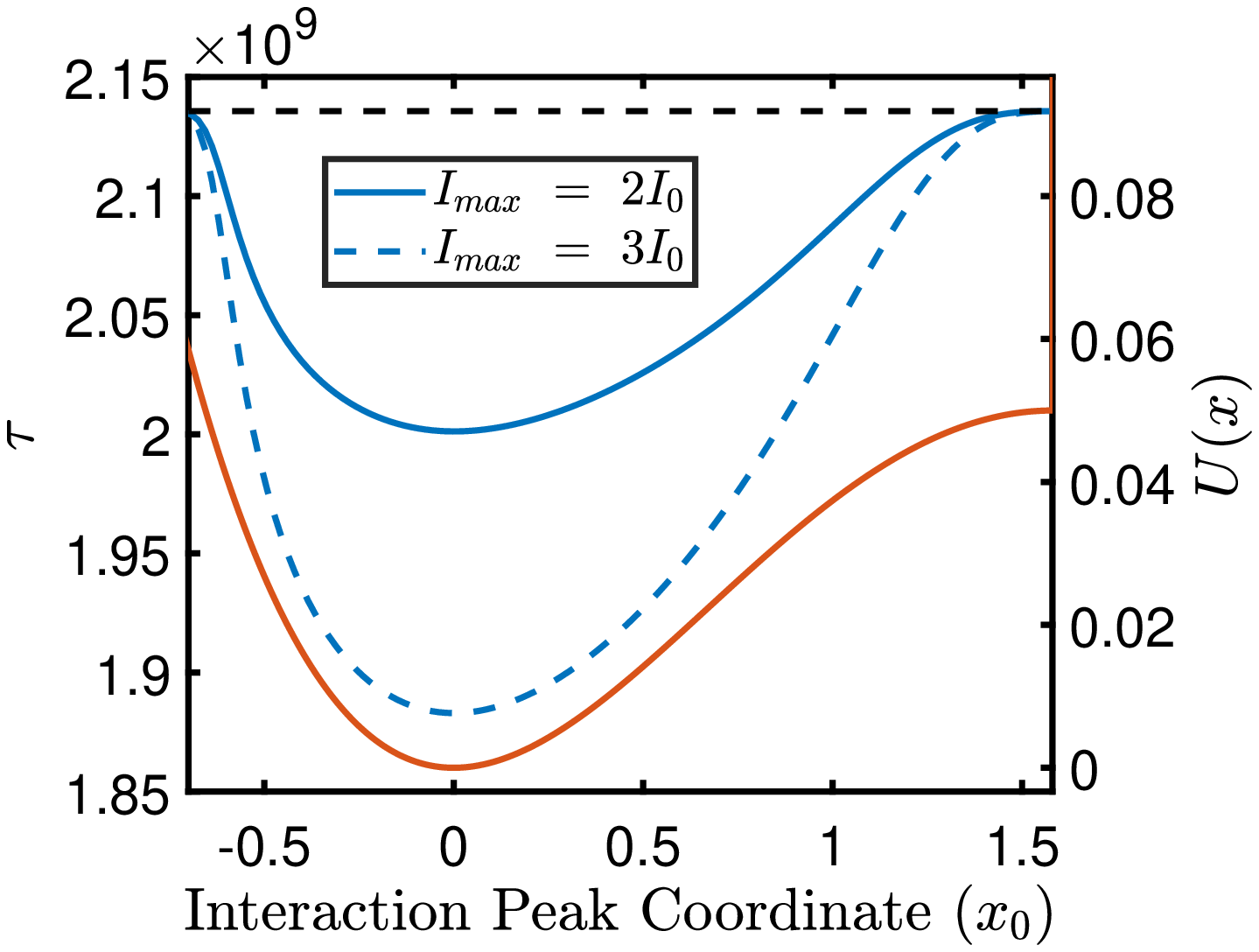}
\caption{}
\label{interaction_underdamped}
\end{subfigure}

\caption{The calculated mean first-passage time as a function of the interaction peak's position along the reaction coordinate, for the (a) overdamped and (b) underdamped regime ($m_e = 1000$a.u). The black line denotes $\tau$ in the absence of an applied blowtorch. Parameters: $D_0 = 0.01$, $\xi_0 = 1$,  $\sigma = 0.05$.}
\label{interaction_data}
\end{figure}

In the underdamped regime, we observe that the largest reduction to $\tau$ occurs when the interaction peak is placed at the minimum. The decrease in $\tau$ agrees with the homogeneous-case solution, with the distinction that reaction coordinates at higher energies in the potential have diminishing contributions to the decreasing $\tau$.
In the overdamped regime, it is seen that the interaction peak results in an increase to $\tau$ as also predicted in the homogeneous case. However, we observe that this is dominated by the increased interaction strength near to the maximum of the potential, while changing the interaction strength in the rest of the potential has negligible effect. This demonstrates that $\tau$ has little regard for the interaction strength in the overdamped regime, except in the region approaching the maximum.

This general analysis arms us with the required physical intution before proceeding to the next section, in which we first observe how a Landauer blowtorch emerges naturally from a simple molecular junction model, then demonstrate the effect on hypothetical chemical reaction rates.

\subsection{Application to a molecule with a single current-carrying molecular orbital}
\label{single_results}

In this section, we analyze the calculated mean first-passage time $\tau$ for our model of a molecular junction. Contrary to section \ref{blowtorch}, the viscosity  and diffusion coefficients will be computed using nonequilibrium Green's functions according to eqs. (\ref{xix}) and (\ref{Dx}). We consider the case of a single electronic level coupled to the left and right leads under some applied bias voltage. The molecular Hamiltonian is simplified to 

\begin{equation}
H_M = (h_0 + \lambda x)d^{\dag}d,
\end{equation}

where subscript $d^\dag$ and $d^\dag$ denote the creation and annihilation operators  for an electron on the molecular orbital, while all previous matrix quantities are simplified to scalars in this regime. Here, the dependency of $H_M$ on the reaction coordinate acts to shift the electronic level, as scaled by the tuneable parameter $\lambda$. The left and right lead are each at room temperature (0.00095a.u) and are symmetrically coupled to the central electronic state such that our level-width function is given by $\Gamma_L = \Gamma_R = 0.03$.

In the interest of consistency, we again utilize the same quartic to describe our classical nuclear potential for the reaction coordinate, which is now acted on by an additional adiabatic force term computed using nonequilibrium Green's functions according to equation \ref{F}. This has the effect of shifting and shallowing/deepening the reaction potential depending on the parameters chosen. We also allow for the manual shift of the external potential along the reaction coordinate according to some parameter $x_a$:

\begin{equation}
U(x) = -a(x-x_a)^2 + b(x-x_a)^4.
\end{equation}

This means that when $x_a = 0$, the potential minimum (ignoring the effects of $F$) occurs at $x=0$, while a positive $x_a$ shifts the input potential to the right. Any bias voltage is applied symmetrically, such that $\mu_L = -\mu_R = V/2$, where $\mu_L$ and $\mu_R$ are the chemical potentials of the left and right leads.

\begin{figure*}
\begin{subfigure}{0.4\textwidth}
\centering
\includegraphics[width=1\textwidth]{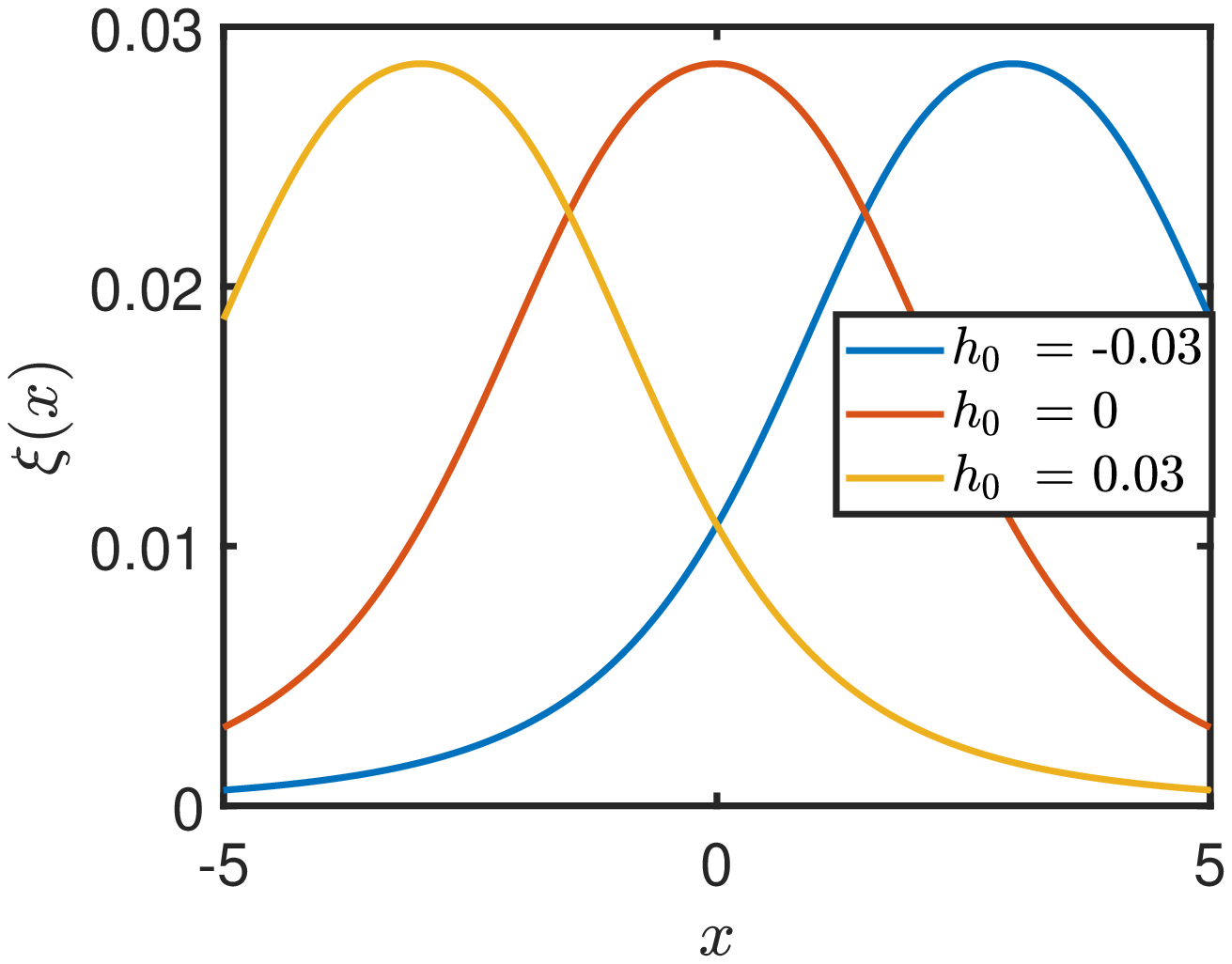}
\caption{}
\label{nona}
\end{subfigure}
\begin{subfigure}{0.4\textwidth}
\centering
\includegraphics[width=1\textwidth]{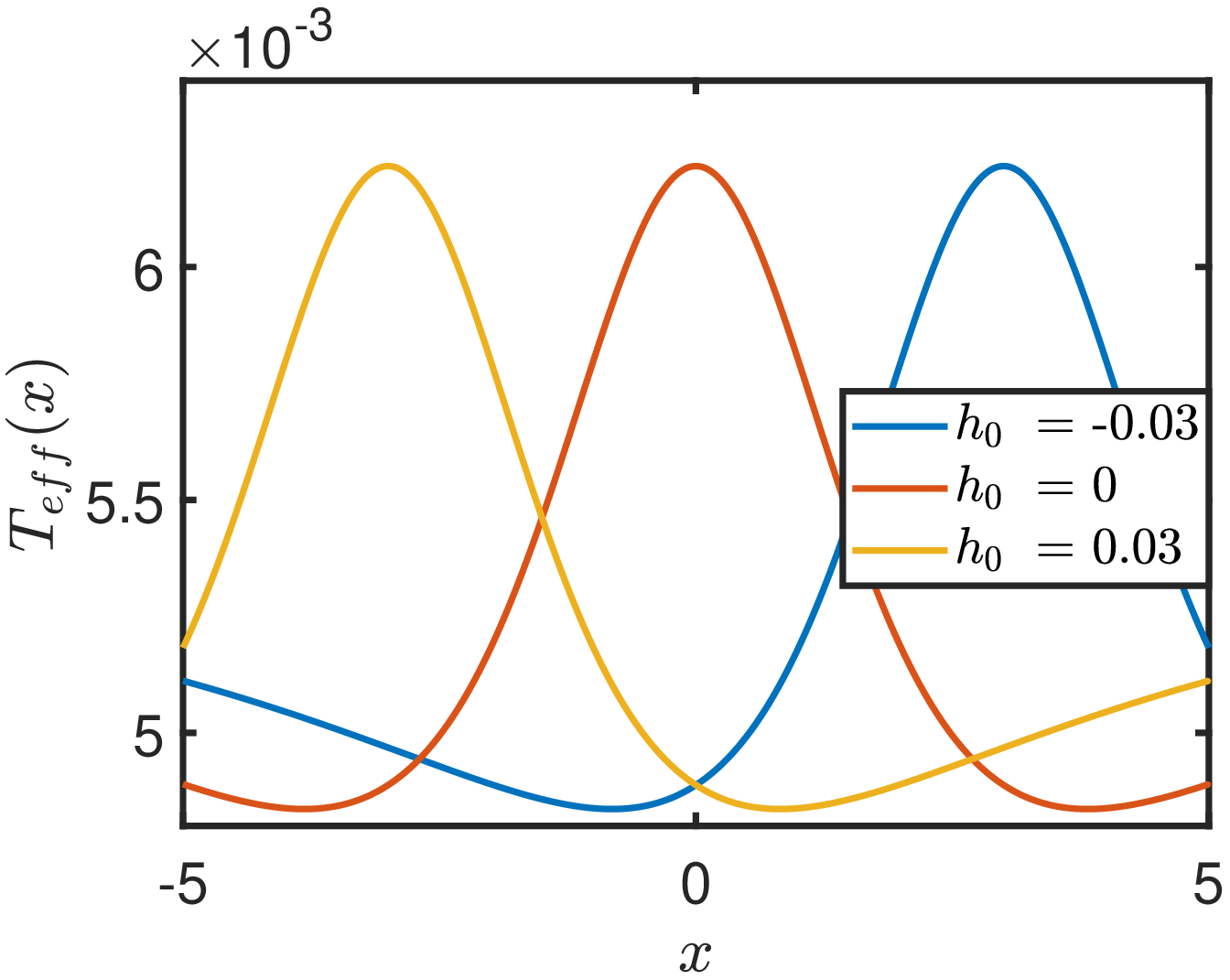}
\caption{}
\label{nonb}
\end{subfigure}
\begin{subfigure}{0.4\textwidth}
\centering
\includegraphics[width=1\textwidth]{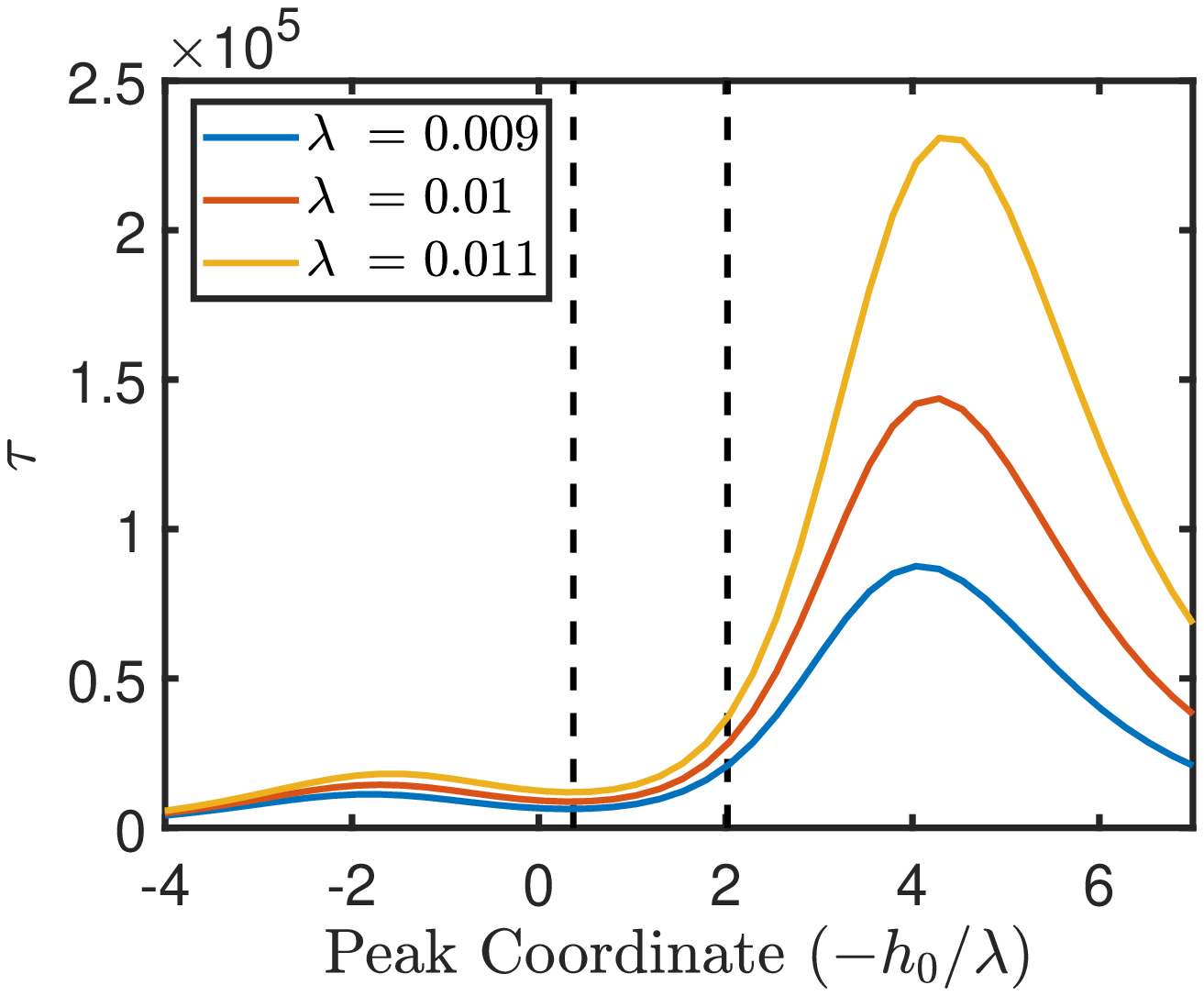}
\caption{}
\label{nonc}
\end{subfigure}
\begin{subfigure}{0.4\textwidth}
\centering
\includegraphics[width=1\textwidth]{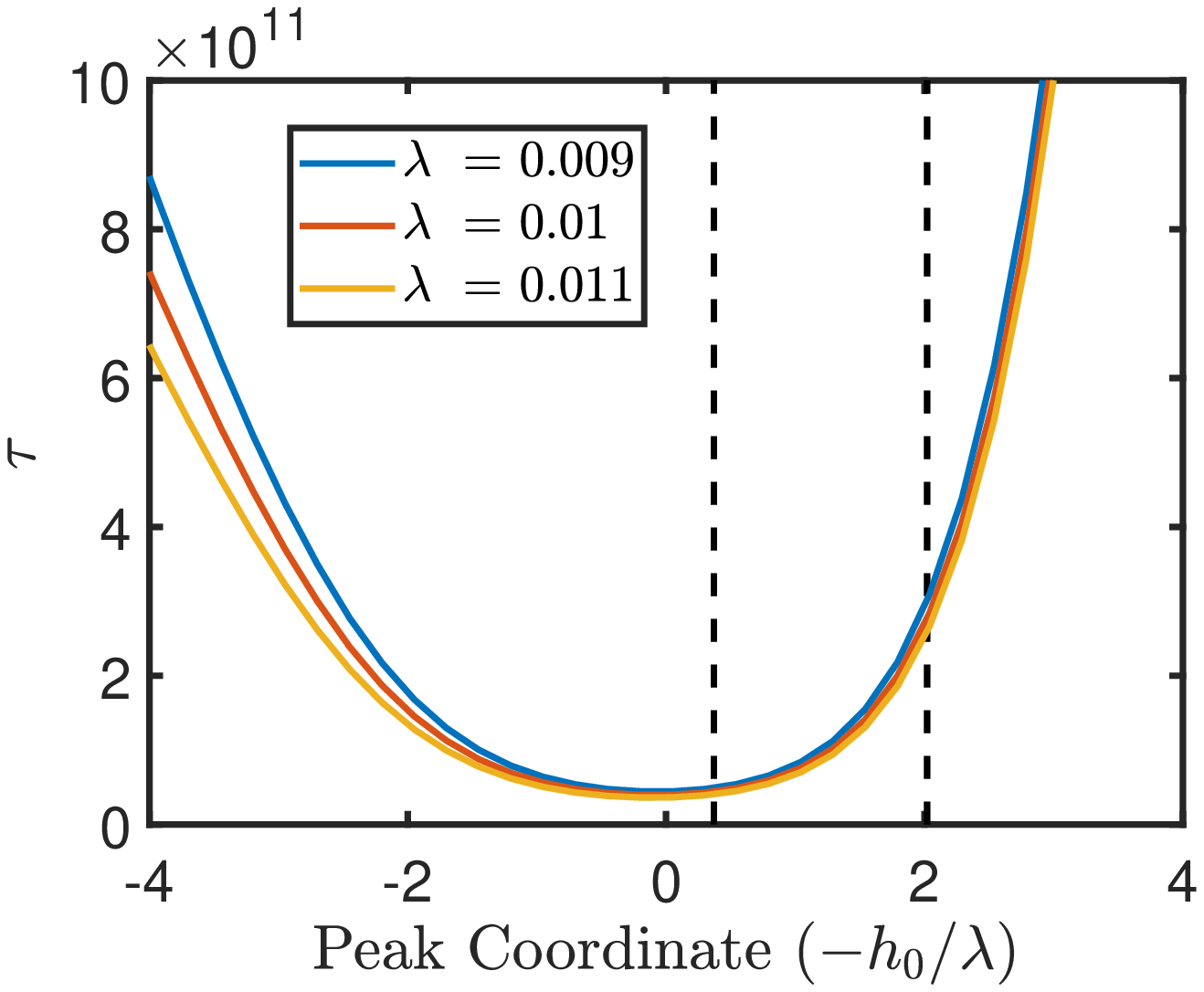}
\caption{}
\label{nond}
\end{subfigure}

\caption{The effect of an applied gate voltage to (a) the viscosity coefficient and (b) the effective temperature. The mean first-passage time $\tau$ in the (c) overdamped and (d) underdamped regime is plotted against the peak coordinate of the viscosity and temperature (as determined by the applied gate voltage) for different $\lambda$. The coordinates of the minimum and maximum of the reaction potential are denoted by the vertical black lines in (c) and (d).}
\end{figure*}

We study the effect of applying a gate voltage to the system, as modelled by a shift in the $h_0$ value. This allows for a degree of controllability of the reaction rates for a given system. Figs.\ref{nona} and \ref{nonb} demonstrate the resultant viscosity coefficient and temperature respectively, as a function of the reaction coordinate. Application of a gate voltage acts to shift the curve along the reaction coordinate. This analysis is performed for a non-zero bias voltage such that the temperature is now inhomogeneous in addition to the viscosity.
In the underdamped regime shown in Fig.\ref{nond}, $\tau$ is minimized when the viscosity and temperature peaks are shifted near to the minimum of the reaction potential (note however, that the minimum in $\tau$ does not occur exactly when the peaks are shifted to the minimum of the potential due to the slight asymmetry of the reaction potential).
In contrast, the overdamped regime displays highly non-trivial behaviour, arising as a result of the interplay between the strength of the viscosity and the temperature. In our analysis of the overdamped regime in the previous section, we noted that the dependence of $\tau$ on the temperature is dominated by the region of steepest ascent up towards the maximum. Here, we again observe this behaviour as the large peak in Fig.\ref{nonc} corresponds to when the dip in the temperature occurs in this region (when the temperature peak has been shifted to the right). A corresponding but smaller peak also occurs due to a shift to the left in the temperature such that the low temperature aligns with the steep region of the potential. The difference in peak sizes arises as a result of the inhomogeneous viscosity, which per the previous section, we know is important in the region near the maximum of the reaction potential. The large peak in $\tau$ occurs when the temperature is low in the steep region, while the viscosity is high towards the maximum. The small peak has a low viscosity near the maximum, explaining its comparatively smaller magnitude. 

\subsection{Model of a two-level molecule}
\label{multi_results}

In this section, we expand the model to consider a two-level system.   The second molecular orbital is not a simple  addition of an extra level  here, since the Green's functions, self-energies and molecular Hamiltonian become $2\times2$ matrices and some nontrivial terms such as the commutator in (\ref{g1A}) will no longer be zero. In our model,
the molecular energy levels is taken to correspond to the bonding and anti-bonding states of a free $H_2^+$ molecule\cite{mcquarrie-qc}. As such, the molecular Hamiltonian now reads

\begin{equation}
H_M(t) = \sum_{ij}h_{ij}(q(t))d^{\dag}_i d_j,
\end{equation}
where {    $d_i^{\dag}$ and $d_j$ } are now in the molecular orbital basis. The electronic Hamiltonian elements are represented in the form of a $2\times 2$ matrix according to 

\begin{equation}
h = 
\begin{pmatrix}
H_b(q) & 0 \\
0 & H_a(q)
\end{pmatrix}
,
\end{equation}
where we use $H_b(q)$ and $H_a(q)$ to denote the bonding and anti-bonding molecular orbitals, respectively, while $q$ is the  bond-length. The values for $H_b(q)$ and $H_a(q)$ are calculated according to molecular orbital theory\cite{mcquarrie-qc} and are given by

\begin{equation}
H_b(q) = \frac{H_{AA} + H_{AB}}{1+S_{AB}},
\end{equation}

and

\begin{equation}
H_a(q) = \frac{H_{AA} - H_{AB}}{1-S_{AB}},
\end{equation}
where $H_{AA}$ and $H_{AB}$ are the Hamiltonian elements in the atomic basis and $S_{AB}$ is the overlap integral between atomic 1s Slater orbitals. The constituent components are then given by

\begin{equation}
H_{AA} = -\frac{1}{2} + e^{-2q}\Big(1+\frac{1}{q}\Big) - \frac{1}{q},
\end{equation}

\begin{figure*}
\begin{subfigure}{0.4\textwidth}
\centering
\includegraphics[width=1\textwidth]{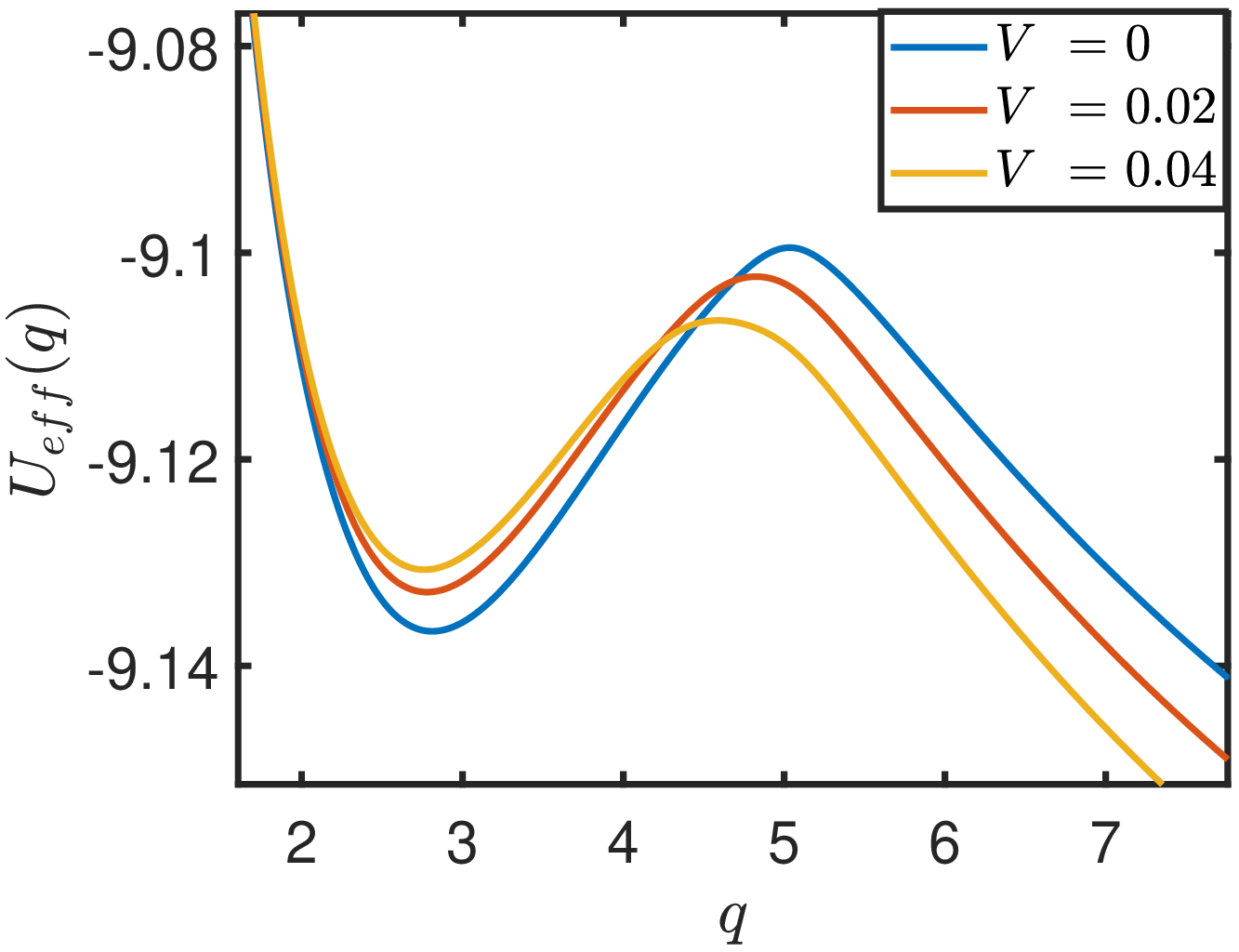}
\caption{}
\label{U_V}
\end{subfigure}
\begin{subfigure}{0.4\textwidth}
\centering
\includegraphics[width=1\textwidth]{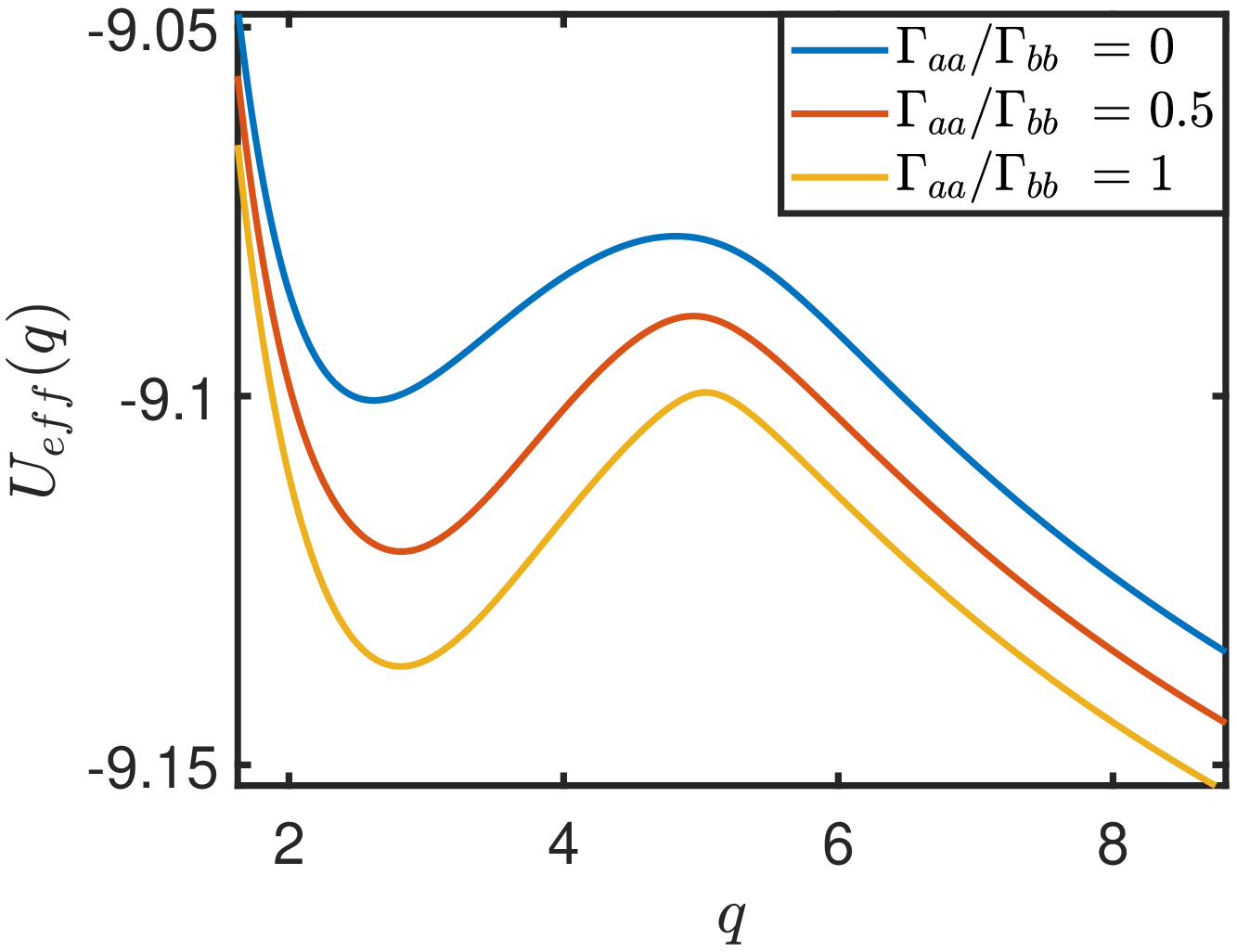}
\caption{}
\label{U_gam}
\end{subfigure}

\caption{The adiabatic potential as a function of the bond-length presented for (a) varying bias voltages and (b) varying the magnitude of leads coupling to $H_a$. Parameters: $V = 0$, $\Gamma_{aa}/\Gamma_{bb} = 1$, unless otherwise specified. $\Gamma_{bb} =0.03$ in all calculations.}
\label{U}
\end{figure*}

\begin{figure*}
\begin{subfigure}{0.4\textwidth}
\centering
\includegraphics[width=1\textwidth]{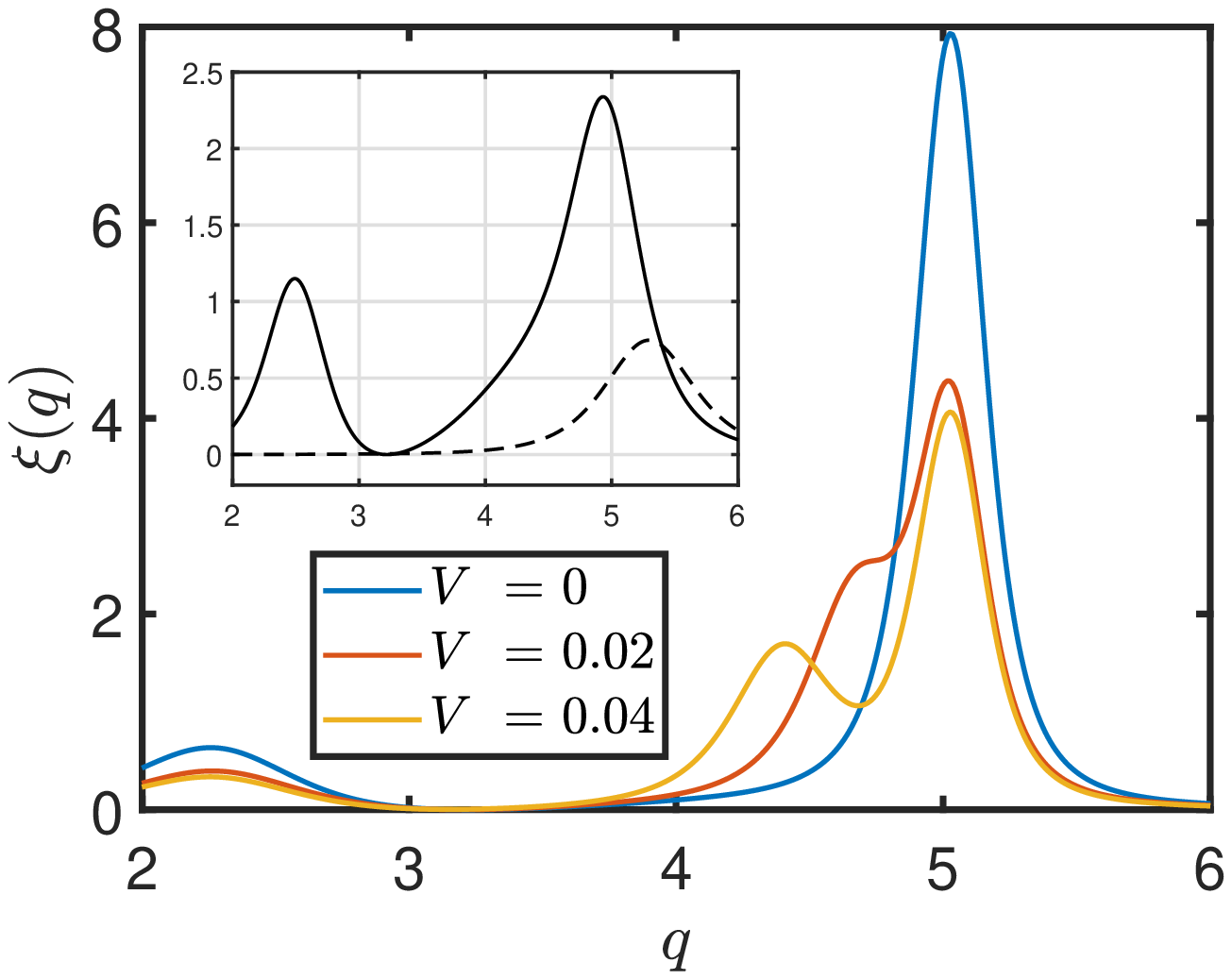}
\caption{}
\label{multi_xi}
\end{subfigure}
\begin{subfigure}{0.4\textwidth}
\centering
\includegraphics[width=1\textwidth]{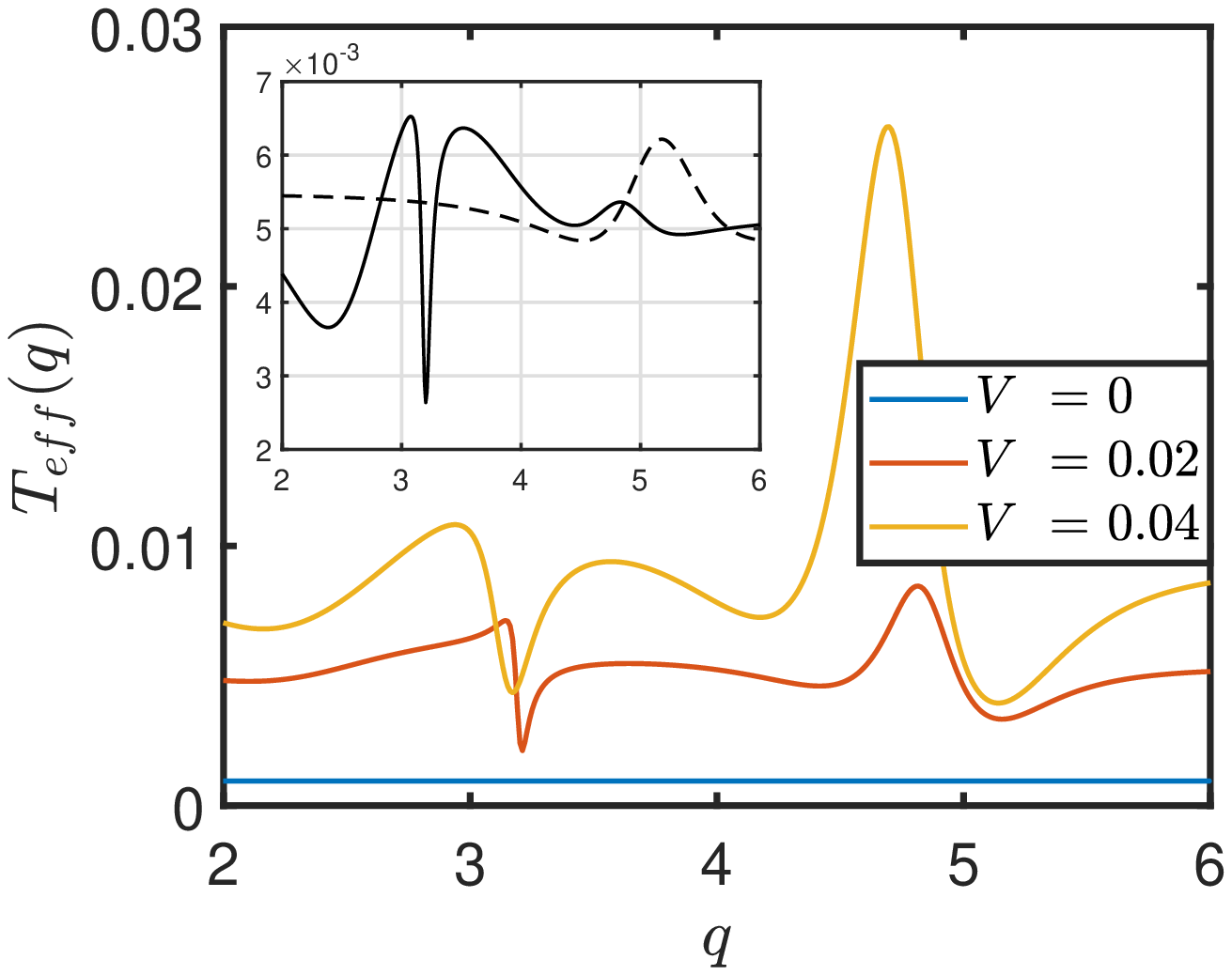}
\caption{}
\label{multi_T}
\end{subfigure}

\caption{{    The effect of varying the bias voltage is shown for the (a) viscosity coefficient and (b) the effective temperature}, as a function of the bond length. Insets: Shows the same quantity at $V=0.02$ for $\Gamma_{aa}/\Gamma_{bb}=0$ (dashed) and $\Gamma_{aa}/\Gamma_{bb} = 0.5$ (solid). Parameters; $\Gamma_{aa}/\Gamma_{bb} = 1$ in the main plots. $\Gamma_{bb} =0.03$ in all calculations.}
\label{multi_lang}
\end{figure*}

\begin{figure*}
\begin{subfigure}{0.4\textwidth}
\centering
\includegraphics[width=1\textwidth]{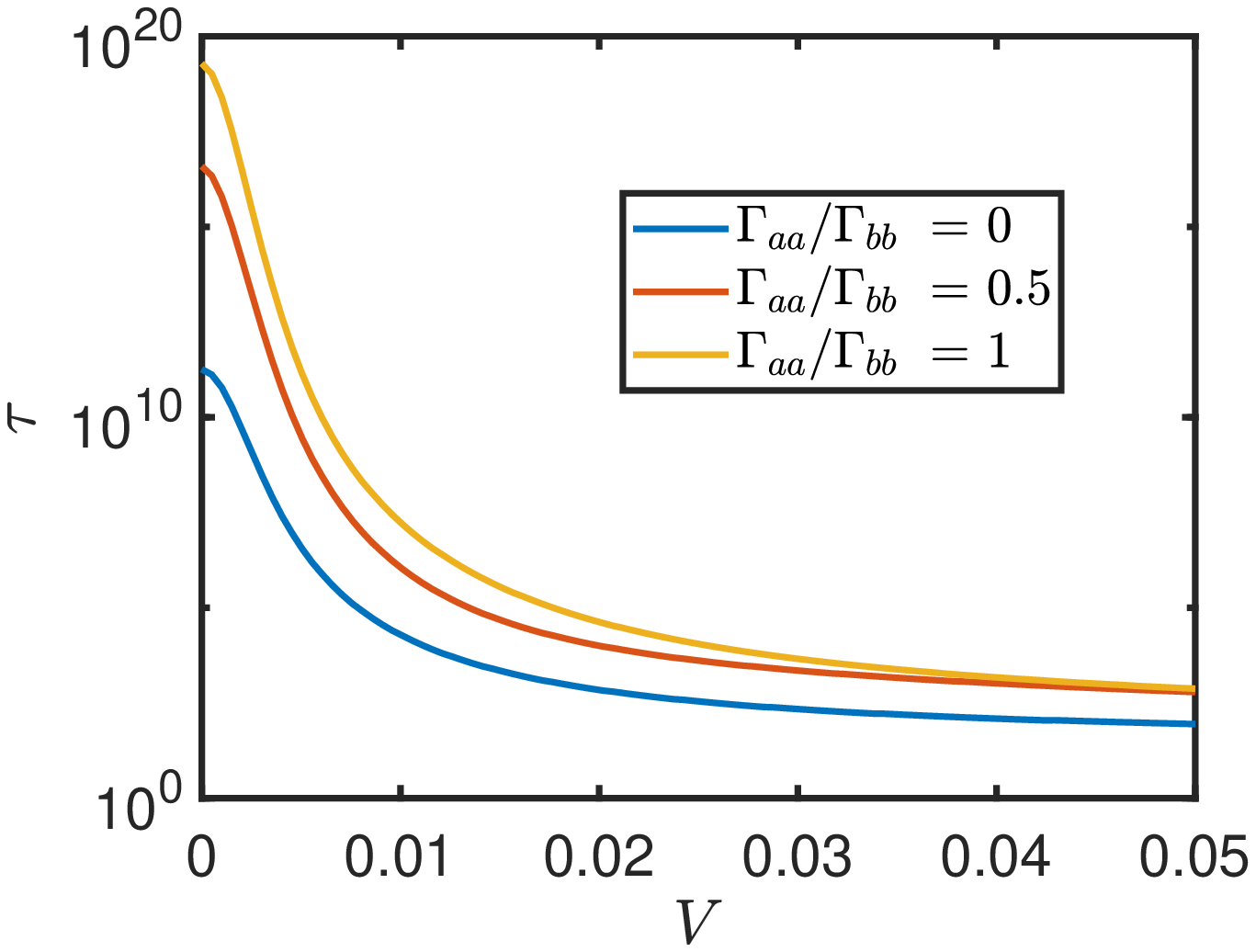}
\caption{}
\label{mul_over}
\end{subfigure}
\begin{subfigure}{0.4\textwidth}
\centering
\includegraphics[width=1\textwidth]{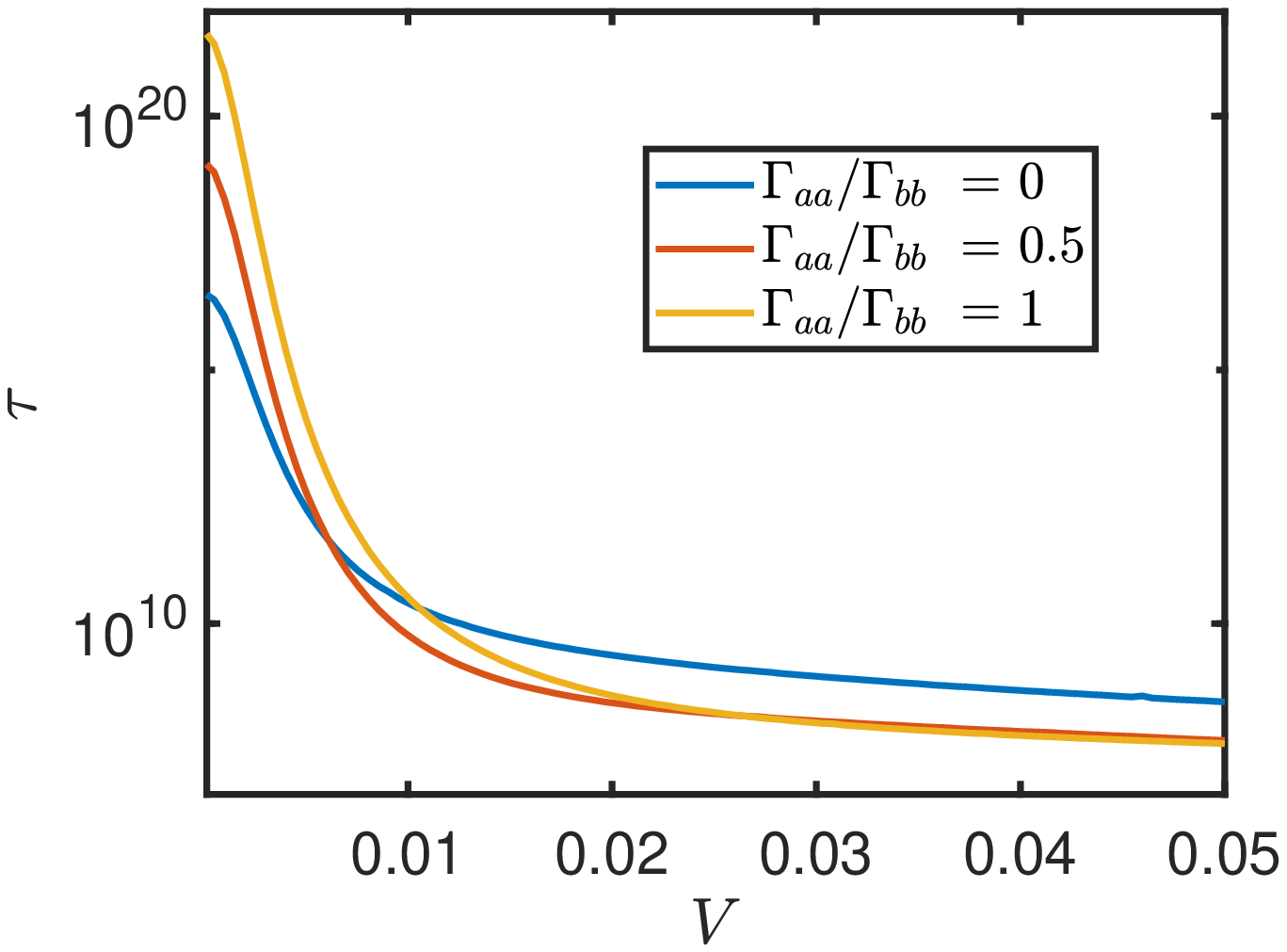}
\caption{}
\label{mul_under}
\end{subfigure}

\caption{The mean first-passage time $\tau$ as a function of the bias voltage, varying the coupling to $H_a$ in the (a) overdamped and (b) underdamped cases.}
\label{multi_MFPT}
\end{figure*}

\begin{equation}
H_{AB} = -\frac{S_{AB}(q)}{2} - e^{-q}(1+q),
\end{equation}

and

\begin{equation}
S_{AB} = e^{-q}(1+q+q^2 / 3).
\end{equation}

In the interest of simplicity, each of the molecular orbitals is symmetrically coupled to the left and right leads; as controlled by the parameter $\Gamma$ which now takes the form of a matrix as per

\begin{equation}
\Gamma_{\alpha} = 
\begin{pmatrix}
\Gamma_{\alpha,bb} & \Gamma_{\alpha,ba} \\
\Gamma_{\alpha,ab} & \Gamma_{\alpha,aa}
\end{pmatrix}
\end{equation}
for the $\alpha$ lead, where the off-diagonal components can be defined according to,

\begin{equation}
\Gamma_{\alpha,ba}=\Gamma_{\alpha,ab} = \sqrt{\Gamma_{\alpha,aa}\Gamma_{\alpha,bb}}.
\end{equation}
In each test, we have $\mu_L = -0.7$ and $\mu_R = \mu_L - V$, while the lead temperatures are again set to room temperature.

 The external potential now represents the classical nuclear repulsion, which in atomic units is given by

\begin{equation}
U(q) = \frac{1}{q}.
\end{equation}

Inclusion of the electronic forces allows us to generate modified electronic potentials for varied parameters in order to assess the molecular stability. Examples of these potentials are shown in Fig.\ref{U}; where in (a) an applied bias voltage is shown to decrease the energy required for bond rupture, while (b) shows the effect of the additional electronic level which when occupied, acts to increase the bond stability.

Along with the shape of the effective potential, the bond stability is also determined by the electronic viscosity and effective temperature, which are demonstrated in Fig.\ref{multi_lang}. In the viscosity coefficient, each curve shows a peak at small $q$, which approximately corresponds to when $H_a$ crosses the fermi-level of the left lead. Likewise, the peaks at large $q$ are a result of $H_b$ crossing the fermi-level of the right, then left, leads (these split peaks merge together when $V=0$). The inset plot demonstrates the effect of allowing an additional transport channel through the excited state which not only introduces the peak at small $q$, but also increases the magnitude of the viscous forces overall. 
The effective temperature is equal to the leads temperature for $V=0$, while non-zero voltages yield a complex array of localized heating and cooling effects, which arise as the energy levels shift in and out of the resonance region as the bond-length is increased.

These competing effects culminate in our calculation of the mean first-passage time, which is demonstrated in Fig.\ref{multi_MFPT} as a function of the bias voltage, for different coupling values to the excited electronic state. In both limiting regimes, an increase to the bias voltage acts to destabilize the bond and decrease $\tau$, both due to the increased effective temperatures and the weakening of the bond due to the current-induced forces. In the overdamped regime, allowing the leads to be coupled to an additional level in the central region has a stabilizing effect for all voltages tested, increasing the average amount of time for bond rupture. The underdamped case shows similar behavior for very low voltages; however, at higher voltages the availability of the additional transport channel through the excited state increases the current-induced forces such that the energy required for a bond-rupture is found more easily, decreasing $\tau$.

\section{CONCLUSIONS}

In this paper, we have demonstrated that {    the rates} of chemical reactions for molecules in electronic junctions depend on
three crucial ingredients; the potential energy surface which defines the energy required for a configuration change or bond rupture, the rate of the energy removal from vibrational to electronic degrees of freedom given by the electronic  viscosity coefficient, and lastly, the effective temperature dynamically established in the molecule.
While the magnitude of these quantities is of high importance, the local distribution of the viscosity and effective temperature along the potential energy surface (Landauer's blowtorch effect) also proves to be critical.

The addition of localized heating and cooling effects as a result of inhomogeneity with respect to the molecular configuration has been shown to induce significant variations in the mean first-passage time, as calculated according to a Fokker-Planck description obtained by separating slow (reaction coordinate) and fast (electronic) time-scales in the Keldysh-Kadanoff-Baym equations for the nonequilibrium Green's functions.  This has been demonstrated for a  single-level molecular junction model, as well as a two-level model inspired by  $H_2^+$ molecular orbitals with the bond length considered as the reaction coordinate. This interplay between the amount of energy required for bond rupture and the energy supplied due to tunneling electrons has been shown to be strongly dependent on the choice of experimentally tuneable parameters for the system. This enables the possibility of a high degree of controllability for molecular junction systems, with promises of controlled initiation of chemical reactions or conversely, enforcing the stability of specific configurations within the system.

\begin{center}
{\bf DATA AVAILABILITY}
\end{center}

The data that supports the findings of this study are available within the article.

\appendix

\section{Derivation of the energy diffusion equation in the underdamped limit \label{Und}}

Following Zwanzig \cite{zwanzig-book}, we can introduce the projection operator 
\begin{equation}
\widehat{O}=\Omega^{-1}(E)\int dxdp\,\,\delta(E-H)\label{P1}
\end{equation}
where 
{   
\begin{equation}
H=\frac{p^{2}}{2m}+U_{\text{eff}}(x)\label{H}
\end{equation}
}
is the Hamiltonian of the bath-free Brownian particle and the microcanonical
partition function is defined as  
\begin{equation}
\Omega(E)=\int dxdp\,\,\delta(E-H)\label{P2}
\end{equation}
(equivalently, $\Omega(E)$ is determined via Eq. (\ref{Om})). Note
that $\widehat{O}$ has an extra factor of $\Omega^{-1}(E)$ in comparison with the operator introduced by Zwanzig.
With our definition, $\widehat{O}^{2}=\widehat{O}$. 
{    Applying the projection operator of Eq. (\ref{P1}) to the Fokker-Planck equation (\ref{FPEa}),  
exploiting 
}
the identity
{   
\[
\widehat{O}\left(-\partial_{x}\frac{p}{m}+\partial_{p}U_{\text{eff}}^{^{\prime}}(x)\right)=0
\]
}
and taking into account that $\xi(x) \ll 1$ (underdamped limit) we
obtain
{   
\begin{equation}
\partial_{t}\widehat{O}\rho(x,p,t)=\widehat{O}\zeta(x)\left\{ \partial_{p}\frac{p}{m}+\partial_{p}^{2}T(x)\right\} \widehat{O}\rho(x,p,t).\label{FPE1}
\end{equation}
}
Changing the differentiation variables
 {   
	($\partial_{p}= m^{-1}\partial_{E}p$) and using the identity $\partial_{p}^{2}\widehat{O}= m^{-1}\partial_{p}p\partial_{E}\widehat{O}$,
}
we obtain the following equation for $\rho(E,t)\equiv\Omega(E)\widehat{O}\rho(t)$ (cf. \cite{zwanzig-book,GK2007}):
\begin{equation}
\partial_{t}\rho(E,t)=\partial_{E}\left\{ \mu(E)+\nu(E)\partial_{E}\right\} \Omega^{-1}(E)\rho(E,t).\label{D1}
\end{equation}
Here $\mu(E)$ and $\nu(E)$ are defined through Eqs. (\ref{mu}) and (\ref{nu}),
and $\rho(E,t)$ is normalized according to $\int dE\rho(E,t)=1$.

\clearpage
\bibliography{rileysrefs}
\end{document}